\documentclass[12pt]{article}

\usepackage{amsmath,amsfonts,amssymb,latexsym,graphicx}

\usepackage{caption}
\usepackage{subcaption}

\def\be{\begin{equation}}
\def\ee{\end{equation}}
\def\bq{\begin{eqnarray}}
\def\eq{\end{eqnarray}}
\def\beq{\begin{eqnarray}}
\def\eeq{\end{eqnarray}}

\def\a{\alpha}
\def\b{\beta}

\begin{document}

\Large
 \begin{center}
\textsc{Sudden Shock Waves in modified gravity}

\hspace{10pt}

\Large
Ignatios Antoniadis$^{1,2,\dagger}$, Spiros Cotsakis$^{3,4,\ddagger}$, Dimitrios Trachilis$^{5,*}$

\hspace{10pt}

\small
$^1$) Laboratoire de Physique Th\'{e}orique et Hautes Energies - LPTHE\\
Sorbonne Universit\'{e}, CNRS 4 Place Jussieu, 75005 Paris, France\\
$^2$) School of Natural Sciences, Institute for Advanced Study\\
Princeton NJ 08540, USA\\
$^\dagger$antoniad@lpthe.jussieu.fr\\
$^3$) Institute of Gravitation and Cosmology\\ RUDN University,
ul. Miklukho-Maklaya 6, Moscow 117198, Russia\\
$^4$) Research Laboratory of Geometry,  Dynamical Systems  and Cosmology\\
University of the Aegean, Karlovassi 83200, Samos, Greece\\
$^\ddagger$skot@aegean.gr\\
$^{5}$) College of Engineering and Technology\\
American University  of the Middle East, Kuwait\\
$^{*}$dimitrios.trachilis@aum.edu.kw
\end{center}

\begin{center}\small June 2023\end{center}
\vspace{3pt}
\normalsize

\noindent \textbf{Abstract.}  We construct a generic asymptotic solution for modified gravity near a sudden singularity. This solution contains a fluid source with no equation of state and is function-counting stable, that is it has eleven independent arbitrary functions of the spatial coordinates as dictated by the Cauchy problem of the theory. We further show that near the sudden singularity the solution has a shock wave character with the same number of free functions in the Jordan and Einstein frame.

\section{Introduction}
Barrow \cite{jdb04a} introduced sudden future singularities in general relativistic cosmology as a counterexample to the expectation that closed Friedmann universes recollapse if the strong energy condition holds \cite{jdb04aa}. Sudden singularities in general relativity are geodesically complete in the smooth \cite{bc13}, or the distributional sense \cite{ker} They are also generally stable with respect to various kinds of small cosmological fluctuations \cite{lip}, and quantum particle production or regularization schemes \cite{fab1,fab2,haro}. Sudden singularities can also be  anisotropic  \cite{tsa}, and have other interesting properties,  e.g., they do not destroy extended objects like cosmic string loops \cite{da}.

Sudden singularities  are in fact of a different nature than most other dynamical singularities encountered in cosmology \cite{iklaoud1,iklaoud2}, in that they can be extended to solutions that have the full number of free functions necessary of a general solution to the field equations \cite{bct}. This property, which we call \emph{function-counting stability}, has an extensive history and was one sought for and extensively studied very early in the history of the subject, cf.  \cite{lk1}-\cite{ll}, and more recently in \cite{kks}-\cite{kks2}, for reviews and more general related issues, see \cite{khalatnikov}-\cite{cy22}.

The extension of the general relativistic, homogeneous and isotropic sudden singularity solution to the case of modified  gravity theories is not automatic: for the case of $f(R)$ gravity it was found in \cite{jdb04b}, while for Brans-Dicke theory the sudden singularity was constructed in \cite{jdbbd}. For the Brans-Dicke theory, we know that the sudden singularity is function-counting stable with respect to generic perturbations \cite{bct1}, but for modified gravity we only know this to be true for any regular solution of the theory \cite{ckt}, and no such result is known for the sudden singularity solution found in \cite{jdb04b}. It is the purpose of this paper to prove the function-counting stability of the $f(R)$-sudden singularity, and also show that the constructed solution has a shock wave character, hence it is `mild' in the sense of Refs. \cite{tip,kr}.

In the next two Sections, we prove function-counting stability in the Jordan and Einstein frame representation of the quadratic $f(R)$ theory, and in Section 4 we establish the Tipler-Krolac conditions and the behaviours of the expansion of the geodesic congruence and its higher-order derivatives that imply the shock wave character of the general solution. Some aspects are briefly given in the last Section. Below we set $8\pi G =1$ and follow the Landau-Lifshitz sign conventions \cite{ll}.

\section{Jordan frame sudden singularity}
We shall consider the $R+\epsilon R^2$ theory in the Jordan frame with a matter source given by a fluid with stress energy tensor
$
T_{j}^{i}=  (\rho +p)u^{i}u_{j}-p\delta _{j}^{i},
$
where the unit 4-velocity of the fluid $u^{i}=(u^{0},u^{\alpha })$ has $u^{0}=u_{0}$, and $u_{i}u^{i}=1$, so that,
$
u_{0}^{2}=1+u_\alpha u^\alpha,
$
 the three arbitrary components $u^\alpha$ of the velocity vector field determine $u^0$. We shall assume that $\epsilon>0$.
Barrow has shown that this theory admits an exact solution with a sudden singularity for any Friedmann model with line element $ds^2=dt^2-a^2 (t)d\Omega^2$, where $t$ is the proper time and $d\Omega^2$ the 3-metric of the flat, open or closed spatial sections. The solution is given by \cite{jdb04b},
\begin{equation}
a(t) = \left(\frac{t}{t_{s}}\right)^{q}\left( a_{s}-1\right) +1-\left(1-\frac{t}{t_{s}}\right)^{n}, \quad a_{s}\equiv a(t_{s}),
\label{sol}
\end{equation}%
with $0<q\leq 1$ and $3<n<4$, $t_s$ is the time of the sudden singularity, $a_s=a(t_s)$,  while the fluid pressure and energy density $p,\rho$ are two unconnected functions, that is there is no assumed equation of state between them. This solution exists in the interval $[0,t_s]$, and there is a big rip singularity at $t=0$ where the expansion rate $H$ diverges, being continuous on $(0,t_s)$, cf. \cite{iklaoud1}.  We note the important fact that geodesics at $t_s$ can be continued past the sudden singularity \cite{bc13}, and this fact makes sudden singularities milder than all other kinds of cosmological  singularities.
At $t_s$, the pressure diverges to plus infinity, but the scale factor $a_s$ is finite (note that $a$ is zero at $t=0$, but $\dot{a}$ blows up there \cite{iklaoud1}), while the higher derivatives of the scale factor diverge as well. A sudden singularity is generically different from a big bang, or a big rip, or even various other types of milder singularities that may develop during the evolution (cf., \cite{iklaoud1}), \cite{iklaoud2} for a fuller discussion and  complete classification of the various types of dynamical singularity that may arise in a cosmological setting).

We assume normal Jordan frame coordinates in a small neighborhood of the sudden singularity, that is a Gaussian normal coordinate system (or `synchronous' coordinates in other terminology \cite{pen72}, \cite{ll}, p. 286),  $(\tilde{t},\tilde{x},\tilde{y},\tilde{z})$, where $(t,x,y,z)$ are standard normal coordinates and $\tilde{t}=(t^2-x^2-y^2-z^2)^{1/2}$ (defined such that the quantity under the square root is positive), and also $\tilde{x}=x/t,\tilde{y}=y/t, \tilde{z}=z/t$, where $\tilde{t}=\textrm{const.}$ being spacelike hypersurfaces orthogonal to the timelike geodesics given by $\tilde{x},\tilde{y},\tilde{z}=\textrm{const.}$, and with $\partial/\partial t$ future pointing. Below we drop the tildes. 
In these coordinates, the spacetime metric has the form, 
\begin{equation}
ds^{2}=dt^{2}-\gamma_{\alpha \beta}dx^{\alpha}dx^{\beta},
\label{met}
\end{equation}
with the spatial metric being $\gamma_{\alpha \beta}=\gamma_{\alpha \beta}(t,x,y,z)$.

We now assume that the metric (\ref{met}) has a future sudden singularity at time $t_s>0$. Following the standard procedure worked out in the references quoted above (e.g., \cite{lk4}, Sect. 4, \cite{bct1}), we seek  a quasi-isotropic  approximation of the metric in the synchronous system in the vicinity of the sudden singularity, keeping only the principal terms in its expansion in powers of $t-t_s$, where the  spatial metric assumes an expansion of the form,
\be
\gamma _{\alpha \beta } = a_{_{\alpha \beta }}+b_{_{\alpha \beta}}(t-t_s)+c_{_{\alpha \beta }}(t-t_s)^2+d_{_{\alpha \beta }}(t-t_s)^3+e_{_{\alpha \beta }}(t-t_s)^{n}+ O((t-t_s)^4),
\label{seriesgab}
\ee
where the coefficients in the formal series are  functions of the space coordinates. We can think of this expansion as being in terms of the new variable $\tau=t-t_s$ (related linearly to the proper time $t$), for which the sudden future singularity has now been shifted to the origin. With this  understanding,   we shall keep the notation $t$ for the proper time below for reasons of simplicity.  Further,  we do not include a term proportional to $t^q$ in the metric expansion (\ref{seriesgab}), in accordance with the defining property of a sudden singularity that for $n\in (N,N+1)$, the $(N+1)$ derivative of the metric diverges while all derivatives of orders less or equal to $N$ are finite (or zero), cf. \cite{jdb04b}.

We shall show below that the expansion (\ref{seriesgab}) represents a  general solution of field equations of the $R+\epsilon R^2$ theory in the neighborhood of  a sudden singularity that contains precisely $11$ arbitrary functions of the spatial coordinates, such a number is required for generality in this case (i.e., function-counting stability).

For the metric (\ref{met}), the $R+\epsilon R^2$ field equations split in the usual way into their $\binom{0}{0}$, $\binom{\beta}{\alpha}$, and $\binom{0}{\alpha }$ components. Using the expansion (\ref{seriesgab}) and after some algebra we find the required asymptotic forms  for the density, pressure, and spatial fluid velocity as follows:
From the $\binom{0}{0}$ component,  we obtain the expansion for the energy density,
\be
\rho=\rho_0+\rho_{n-3} t^{n-3}+\rho_1t+O(t^2),\label{seriesdensity}
\ee
from the trace of the $\binom{\beta}{\alpha}$ component,  we have for the pressure,
\be
p=p_{n-4} t^{n-4}+O(t^0),
\label{seriespressure_n-4}
\ee
while from  the dominant term of the $\binom{0}{\alpha }$ components, we find,
\bq
u_{\alpha}=u_{\alpha,{4-n}}t^{4-n}+O(t).
\label{velocities}
\eq
In the expansions (\ref{seriesdensity}), (\ref{seriespressure_n-4}), and (\ref{velocities}), all coefficients are functions of the basic spatial functions $a,b,c,d,e$ that appear in Eq. (\ref{seriesgab}). The  coefficient $p_{n-4}$ in Eq. (\ref{seriespressure_n-4}) will be useful in Section 4, and so we give it here for convenience,
\be\label{p}
p_{n-4}=2n(n-1)(n-2)(n-3)\epsilon e,
\ee
with $e=\textrm{Tr}e_{\alpha \beta}$. We then find that the only further constraint comes  from Eq. (\ref{seriespressure_n-4}), namely, we obtain,
\begin{equation}
b^{\beta}_{\alpha}=\frac{b}{3}\delta^{\beta}_{\alpha},
\label{series_beta_alpha_result}
\end{equation}
while no other constraint is possible from the remaining field equations. 
Hence, taking into account Eqns. (\ref{seriesdensity}), (\ref{seriespressure_n-4}), (\ref{velocities}), and (\ref{series_beta_alpha_result}), we find $6+1+6+1+1=15$ independent functions from the initial data: $6+1+6$ come from  $a_{\alpha \beta}, b_{\alpha \beta}$, and $c_{\alpha \beta}$, and the remaining two come about because the coefficients $d_{\alpha \beta}$, and $e_{\alpha \beta}$ appear in the expansion coefficients only through their traces $d,e$\footnote{The reason for  this is the fact that in the $f(R)$ field equations, the symbols $d_{\alpha\beta},e_{\alpha\beta}$ enter only in the higher order terms. These terms are proportional to $\nabla\nabla R $ and $\Box R$, with $R$  the scalar curvature,  and so only the \emph{trace} of the Ricci curvarure enters in these terms and no other of its components.}, while $p, \rho,u_{\alpha}$ are fully determined (no other free functions in them).
Subtracting the $4$ coordinate covariances which may still be used to remove four functions, leaves $11$ independent arbitrary functions of the three space coordinates on a surface of constant $t$ time. This is exactly the maximal number of independent arbitrary spatial functions expected in a local representation of part of the general solution of the quadratic equations near a sudden singularity. Hence, the solution (\ref{met})-(\ref{seriesgab}) is function-counting stable.

\section{Conformal frame sudden singularity}
The corresponding metric in the conformal frame is given by,
\begin{equation}
\Tilde{g}_{ij}=\Omega^2 g_{ij},\quad\Omega^2 = f'(R)=1+2\epsilon R = e^{\phi}.
\label{gtilde}
\end{equation}
Since in Jordan frame $g_{00}=1$ and $g_{0 \alpha}=0$, we have,
\begin{eqnarray}
\Tilde{g}_{00} &=& \Omega^2 = e^{\phi}=(e^{\phi})_{0} + (e^{\phi})_{1}t + (e^{\phi})_{n-2}t^{n-2} + O(t^2),
\label{tildeg00}\\
\Tilde{g}_{0\alpha} &=& 0,
\label{tildeg0a}\\
\Tilde{\gamma}_{\alpha \beta} &=&-\Tilde{g}_{\alpha \beta} = \Tilde{a}_{\alpha \beta} + \Tilde{b}_{\alpha \beta}t + \Tilde{c}_{\alpha \beta}t^{n-2} + O(t^2).
\label{tildegab}
\end{eqnarray}
Also, differentiating with respect to $t$ in (\ref{tildeg00}), we obtain,
\begin{equation}
e^{\phi}\partial_t{\phi}=(e^{\phi})_{1}+(n-2)(e^{\phi})_{n-2}t^{n-3}+O(t),
\label{derivephi}
\end{equation}
thus,
\begin{eqnarray}
{\phi} &=& {\phi}_{0} + \frac{(e^{\phi})_{1}}{(e^{\phi})_{0}}t + \frac{(e^{\phi})_{n-2}}{(e^{\phi})_{0}}t^{n-2} + O(t^2) \nonumber \\
&=& {\phi}_{0} + {\phi}_{1}t + {\phi}_{n-2}t^{n-2} + O(t^2).
\label{phiseries}
\end{eqnarray}
All coefficients are functions of the spatial coordinates.
(We note that if we set $(e^{\phi})_{0}=1$ in (\ref{tildeg00}), we would gauge-fix the lapse to be equal to one also conformally, something we choose not to do in the following.) Then the conformal frame field equations are,
\begin{eqnarray}
\Tilde{G}^i_j &=& \Tilde{R}^i_j-\frac{1}{2}\delta^i_j \Tilde{R}={T^i_j}^{(M)}+{T^i_j}^{(\phi)},
\label{Einstein}  \\
\Box_{\Tilde{g}}\phi &=& \frac{dV}{d\phi},
\label{wave}
\end{eqnarray}
where,
\begin{eqnarray}
{T^i_j}^{(M)} &=& (\Tilde{\rho}+\Tilde{p})\Tilde{u}^{i}\Tilde{u}_{j}-\Tilde{p}{\delta}^{i}_{j},
\label{matter}  \\
{T^i_j}^{(\phi)} &=& \partial^i{\phi}\partial_j{\phi}-\frac{1}{2}{\delta}^{i}_{j}\Box_{\Tilde{g}}\phi-{\delta}^i_j V(\phi),
\label{en_mom}
\end{eqnarray}
and the potential expansion is found to be,
\begin{eqnarray}
V(\phi) &=& \frac{1}{2}[f'(R)]^{-2}[Rf'(R)-f(R)] \nonumber \\
&=& \frac{1}{2}(1+2\epsilon R)^{-2}\epsilon R^2 \nonumber \\
&=& \frac{1}{8\epsilon}(1-e^{-\phi})^{2} \nonumber \\
&=& \big(V(\phi)\big)_{0} + \big(V(\phi)\big)_{1}t + \big(V(\phi)\big)_{n-2}t^{n-2} + O(t^2).
\label{V(phi)}
\end{eqnarray}
The coefficients in the expansions of the density, pressure and spatial velocities are not so complicated as those in the Jordan frame, and so we can give them here directly. From the $\binom{0}{0}$ component of (\ref{Einstein}), we have,
\begin{equation}
\Tilde{\rho} = \Tilde{R}^{0}_{0}-\frac{1}{2}\Tilde{R}-[e^{-\phi}(\partial_{t}\phi)^2 -\frac{1}{2}{\Box}_{\Tilde{g}}\phi -V(\phi)],
\label{rhoconformal}
\end{equation}
thus,
\begin{eqnarray}
\Tilde{\rho} &=& \frac{1}{2}\Tilde{P}_{0}+\frac{1}{8}\Tilde{b}^{2}-\frac{1}{8}\Tilde{b}^{\beta}_{\alpha}\Tilde{b}^{\alpha}_{\beta} \nonumber \\
&-& \left \{2(e^{-\phi})_{0}{\phi}_{2}-{\phi}_{2}-\frac{1}{2}\Tilde{a}^{\alpha \beta}\big[\partial_{\alpha}(\partial_{\beta}{\phi}_{0})-\frac{1}{2}\Tilde{b}_{\alpha \beta}{\phi}_{1}-(\Tilde{\Gamma}^{\gamma}_{\alpha \beta})_{0}\partial_{\gamma}{\phi}_{0}\big]-\big (V(\phi) \big )_{0}\right \} \nonumber \\
&+& O(t^{n-3}).
\label{rhoconformalseries}
\end{eqnarray}
From the $\binom{\beta}{\alpha}$ component of (\ref{Einstein}), we get,
\begin{equation}
\Tilde{p} \delta^{\beta}_{\alpha} = -\Tilde{R}^{\beta}_{\alpha}+\frac{1}{2}\Tilde{R}\delta^{\beta}_{\alpha}+[-\Tilde{\gamma}^{\beta \gamma}\partial_{\gamma}\phi \partial_{\alpha}\phi -\frac{1}{2}\delta^{\beta}_{\alpha}{\Box}_{\Tilde{g}}\phi -\delta^{\beta}_{\alpha}V(\phi)],
\label{pconformal}
\end{equation}
so the trace gives,
\begin{equation}
\Tilde{p} = \big [-\frac{1}{3}(n-2)(n-3)\Tilde{c} -
4\pi G (n-2)(n-3)\frac{(e^{\phi})_{n-2}}{[(e^{\phi})_{0}]^{2}} \big ]t^{n-4} + O(t).
\label{pconformalseries}
\end{equation}
Using (\ref{pconformal}) and (\ref{pconformalseries}), the $(n-4)$-order terms are,
\begin{equation}
-\frac{1}{3}(n-2)(n-3)\Tilde{c}\delta^{\beta}_{\alpha} = \frac{1}{2}(n-2)(n-3)\Tilde{c}^{\beta}_{\alpha}
-\frac{1}{2}(n-2)(n-3)\Tilde{c}\delta^{\beta}_{\alpha},
\label{tildec}
\end{equation}
and so we find that,
\begin{equation}
\Tilde{c}^{\beta}_{\alpha}=\frac{\Tilde{c}}{3}\delta^{\beta}_{\alpha}.
\label{relationtildec}
\end{equation}
(This last relation for $\Tilde{c}_{\alpha \beta}$ was also found in Refs. \cite{bct}, \cite{bct1}.)
From the $\binom{0}{\alpha}$ component of (\ref{Einstein}), we get,
\begin{equation}
(\Tilde{\rho}+\Tilde{p})\Tilde{u}_{\alpha} = \Tilde{R}^{0}_{\alpha} - e^{-\phi}\partial_{t}\phi \partial_{\alpha}\phi.
\label{uaconformal}
\end{equation}
Taking into account (\ref{rhoconformal}) and (\ref{pconformal}), we find,
\begin{equation}
\Tilde{u}_{\alpha} = \frac{3}{(n-2)(n-3)\Tilde{c}} \big[-\frac{1}{2}(\nabla_{\beta}\Tilde{b}^{\beta}_{\alpha}-\nabla_{\alpha}\Tilde{b})
- (e^{-\phi})_{0}{\phi}_{1}\partial_{\alpha}{\phi}_{0}\big]t^{4-n}.
\label{tildeuaseries}
\end{equation}
Finally, from (\ref{wave}), we find that,
\begin{equation}
e^{\phi}\nabla_{0}({\partial_t}\phi) +e^{2\phi}\Tilde{g}^{\alpha \beta}\nabla_{\alpha}(\partial_{\beta}\phi) = \frac{1}{4\epsilon}(e^{\phi}-1),
\label{wave2}
\end{equation}
and for the $(n-4)$-order term, we obtain,
\begin{equation}
(n-2)(n-3)(e^{\phi})_{0}\frac{(e^{\phi})_{n-2}}{(e^{\phi})_{0}} = 0,
\label{wave3}
\end{equation}
so that,
\begin{equation}
(e^{\phi})_{n-2} = {\phi}_{n-2} = 0.
\label{relationphi}
\end{equation}
We note that the conservation laws of the conformal equations following from (\ref{matter}) and (\ref{en_mom}) do not give any additional constraint on the initial data.

In the conformal frame, we may therefore count as follows. We expect $6 \times \Tilde{g}_{\alpha \beta}$, and
$6 \times \dot{\Tilde{g}}_{\alpha \beta}$, plus $5$ from $\Tilde{u}_{\alpha}, \Tilde{p}$ and $\Tilde{\rho}$, and $2$ additional from $\phi$ and $\dot{\phi}$ giving a total of $19$ independent functions. Subtracting $4$ constraints and $4$ by the coordinate covariances, we are left with $11$ free functions for general solution.
From Eqns. (\ref{rhoconformalseries}), (\ref{pconformalseries}), (\ref{relationtildec}),  (\ref{tildeuaseries}), and (\ref{relationphi}), we get $6+6+1=13$ free functions from
$(\Tilde{a}_{\alpha \beta}, \Tilde{b}_{\alpha \beta}, \Tilde{c}_{\alpha \beta})$, plus $1+1+0=2$ free functions from $({\phi}_{0}, {\phi}_{1}, {\phi}_{n-2})$.
Subtracting the $4$ coordinate covariances in our found solution, we see that there are $15-4=11$ independent functions of the three space coordinates. This proves function-counting stability in the conformal frame with the same number of free functions as that required for stability on the Jordan frame, namely 11 functions.

\section{Shock waves}
We shall define the extrinsic curvature in synchronous coordinates in the standard way, $K_{\a\b}=\partial\gamma_{\a\b}/\partial t$, and denote  its trace by $K=\textrm{Tr} K_{\a\b}$. For a geodesic congruence, $K$ is the expansion of the congruence, that is the fractional rate of change in its cross-sectional volume. Near the singularities predicted by the singularity theorems in general relativity, $K\rightarrow -\infty$, that is the congruence collapses to zero volume and a conjugate point forms showing that spacetime itself collapses.

However, at the moment of a sudden singularity, we have $K\rightarrow b,b>0, $ instead of collapse, which means that  the congruence passes without intersecting through a minimum nonzero volume determined by $b$. The hypersurface containing this volume has a discontinuity \emph{only} in the second derivatives of the spatial metric, that is $\dot{K}\rightarrow \infty$. This is called a shock wave, a situation with singularities appearing  only in the second derivatives of the metric. This effect has been shown to occur in general relativity \cite{bc13}, as well as in Brans-Dicke theory \cite{bct1}.

Below we show that the sudden singularity constructed in this paper also has a shock wave character like the two cases mentioned above. In fact, the present situation is even milder than those in general relativity and Brans-Dicke theory as we shall now show: the discontinuities in the metric appear only in the third derivatives of the metric.

The asymptotic behaviours of the Jordan frame components of the Ricci tensor on approach to the sudden singularity are, $R_{00} \sim t^0$, $R_{0\alpha} \sim t^0$, and $R_{\alpha \beta} \sim t^0$, and also $u^0 \sim t^0$, $u^{\alpha} \sim t^{4-n}$.
Hence, the Tipler \cite{tip}  and Krolak \cite{kr} conditions follow, while using the Ricci tensor and velocity asymptotics, the strong energy condition $R_{ij}u^{i}u^{j} \geq 0$ gives directly,
\begin{equation}
\frac{1}{4}b^{\beta}_{\alpha}b^{\alpha}_{\beta}-c \geq 0,
\label{sec1}
\end{equation}
or, using (\ref{series_beta_alpha_result}), we find that,
\begin{equation}
c \leq \frac{1}{8}b^2.
\label{sec2}
\end{equation}
In quadratic theory, the strong energy condition becomes,
\begin{equation}
(T_{ij}-\frac{1}{2}g_{ij}T)u^{i}u^{j} \geq 0,
\end{equation}
and therefore since the pressure must diverge to $+\infty$ at the sudden singularity \cite{jdb04b}, using  Eq. (\ref{p}) we find that,
\begin{equation}
\epsilon e \geq 0
\end{equation}
Using the field equations in the Jordan frame, we obtain the following asymptotics for the extrinsic curvature and its derivatives:
\be\label{3dotK}
K \rightarrow b,\,\,
\dot{K} \rightarrow 2c-\frac{1}{3}b^2 <0, \,\,
\ddot{K} \rightarrow 6d-2bc +\frac{2}{9}b^3, \,\,
\dddot{K} \rightarrow \infty.
\ee
These results imply  the (generalized) shock wave nature of the general solution in both Jordan and conformal frames.

It is easy to see that the shock wave behaviour is also present in the  conformal frame. The asymptotic behaviours of the components of the Ricci tensor in the conformal frame frame on approach to the sudden singularity are, $\Tilde{R}_{00} \sim t^{n-4}$, $\Tilde{R}_{0\alpha} \sim t^0$, and $\Tilde{R}_{\alpha \beta} \sim t^{n-4}$, and also $\Tilde{u}^0 \sim t^0$, $\Tilde{u}^{\alpha} \sim t^{4-n}$. The strong energy condition $\Tilde{R}_{ij}\Tilde{u}^{i}\Tilde{u}^{j} \geq 0$ then  gives,
\begin{equation}
-\frac{1}{2}(n-2)(n-3)\Tilde{c}t^{n-4} \geq 0,
\label{sec11}
\end{equation}
that is,
\begin{equation}
\Tilde{c}<0.
\label{sec22}
\end{equation}
For the extrinsic curvature and its derivative we get,
\be
\Tilde{K} \sim\Tilde{b}+(n-2)\Tilde{c}t^{n-3},\,\,
\dot{\Tilde{K}} \sim (n-2)(n-3)\Tilde{c}t^{n-4},
\ee
and therefore we find that $\Tilde{K}\rightarrow\Tilde{b},\,\dot{\Tilde{K}} \rightarrow -\infty$,
which means that the shock wave behaviour is also a feature of the conformal frame.

We therefore find that the shock wave behaviour is invariant under the the conformal transformation and occurs in both conformally related frames.

\section{Discussion}
In this paper we studied the nature of generic cosmological solutions near the sudden singularity in the $R+\epsilon R^2$ modified gravity with a fluid source having unconnected $p$ and $\rho$ in the Jordan and Einstein frames.  We showed that such solutions exist and are general in the sense of  having $11$ free functions, that is equal to the required number for a general solution of the theory. This solution admits a shock wave interpretation, that is near the sudden singularity is very mild, further to being geodesically complete (as can be seen by an argument similar to that in  \cite{bc13}).

Since the generic form (\ref{seriesgab}) of the `sudden' metric in the asymptotic region is the same for an smooth $f(R)$ theory (cf. \cite{jdb04b}, discussion after Eq. (14)), we expect our results to continue to hold for any $f(R)$ theory such that the conformal potential $V(\phi)$ at the origin satisfies $V''(0)>0$ (that is when spacetime is flat). Therefore it appears  that the sudden metric  (\ref{sol}) is  function-counting stable  in all these frameworks, whereas in general such a conclusion is not true for other vacuum or fluid-filled solutions. We believe that this result adds to the physical plausibility of the sudden singularity solutions.

\section*{Acknowledgments}
We are grateful to an anonymous referee for constructive comments. The research of S.C.  was funded by RUDN university  scientific project number FSSF-2023-0003.


\begin{thebibliography}{99}
\bibitem{jdb04a}J. D. Barrow,	Class. Quantum Grav. 21, L79 (2004)
\bibitem{jdb04aa}J. D. Barrow, G. J. Galloway and F. J. Tipler, Mon.
Not. R. Astron. Soc. 223 (1986) 835
\bibitem{bc13}J. D. Barrow and S. Cotsakis, Phys. Rev. D88 (2013) 067301; arXiv:1307.5005
\bibitem{ker} Z. Keresztes, L. A. Gergely and A. Yu. Kamenshchik, Phys. Rev. D86 (2012) 063522
\bibitem{lip} J.D. Barrow and S.Z.W Lip, Phys. Rev. D80
(2009) 043518

\bibitem{fab1} J.D. Barrow, A.B. Batista, J.C. Fabris and S. Houndjo,
Phys. Rev. D 78 (2008) 123508
\bibitem{fab2}J.D. Barrow, A.B. Batista, J.C. Fabris and S. Houndjo, Phys. Rev. D84 (2011) 123518
\bibitem{haro}J. de Haro, J. Amoros and E. Elizalde, Phys. Rev. D85 (2012) 123527
\bibitem{tsa}J.D. Barrow and C.G. Tsagas, Class. Quantum Grav. 22 (2005) 1563
\bibitem{da}A.Balcerzak and M. Dabrowski, Phys. Rev. D73 (2006) 101301

\bibitem{iklaoud1}S. Cotsakis and I. Klaoudatou, J. Geom. Phys. 55
(2005) 306--315; arXiv:gr-qc/0409022
\bibitem{iklaoud2}S. Cotsakis  and I. Klaoudatou, J. Geom. Phys. 57
(2007) 1303-1312; arXiv:gr-qc/0604029

\bibitem{bct} J. D. Barrow, S. Cotsakis and A. Tsokaros,  Class.
Quantum Grav. 27 (2010) 165017

\bibitem{lk1}E. M. Lifshitz and I. M. Khalatnikov, Sov. Phys. JETP
12 (1961) 558

\bibitem{lk2}E. M. Lifshitz and I. M. Khalatnikov, Sov. Phys. JETP
12 (1961) 108

\bibitem{lk3}E. M. Lifshitz, V. V. Sudakov, and I. M. Khalatnikov, Sov. Phys. JETP
13 (1961) 1298

\bibitem{lk4}E. M. Lifshitz and I. M. Khalatnikov, Adv. Phys. 12 (1963) 185.
\bibitem{ll} L.  Landau  and  E. M. Lifshitz, \textit{The Classical Theory of
Fields}, 4th rev. ed. (Oxford, Pergamon, 1975)


\bibitem{kks} Khalatnikov I M, A Y Kamenshchik, and A A Starobinsky, Class. Quantum Grav. 19 (2002) 3845
\bibitem{kkms}Khalatnikov I M, A Y Kamenshchik, M Martellini and A A Starobinsky 2003 J. Cosmol. Astropart. Phys. 03 (2003) 001

\bibitem{ckt}S. Cotsakis, S. Kadry,  D. Trachilis, Int. J. Mod. Phys. A 31 (2016) 1650130; arXiv:1212.2412

\bibitem{kks2} Khalatnikov I M, A Y Kamenshchik, and A A Starobinsky, 	JETP 129, No. 4, 486-494 (2019) [ZhETF 156, No. 4, 581-584 (2019)]; arXiv:1312.0237 [astro-ph.CO]

\bibitem{khalatnikov}I.M. Khalatnikov, A.Yu. Kamenshchik, Phys.Usp. 51 (2008) 609-616; e-Print: 0803.2684 [gr-qc]
\bibitem{jdb14}J.D. Barrow,  Phys. Rev. D89, 064022 (2014); arXiv:1401.4344 
\bibitem{jdb17}J.D. Barrow, \emph{Some generalities about generality},  In:  'The Philosophy of Cosmology, eds. K. Chamcham, J. Silk, J.D. Barrow \& S. Saunders, (Cambridge UP, 2017)
\bibitem{cy22}S. Cotsakis and A. P. Yefremov, Phil. Trans. R. Soc. A380 (2022) 20210191; DOI: 10.1098/rsta.2021.0191, arXiv:2203.16443
    
\bibitem{jdb04b}J.D. Barrow,  Class. Quantum Grav. 21 (2004) 5619

\bibitem{jdbbd}J. D. Barrow,	Class. Quantum Grav. 37 (2020) 065014; arXiv:1909.09519

\bibitem{bct1} J. D. Barrow, S. Cotsakis, D. Trachilis, Eur. Phys. J. C 80 (2020) 1197

\bibitem{tip} F. J. Tipler, Phys. Lett. A 64 (1977) 8
\bibitem{kr} A. Krolak Class. Quantum Grav. \textbf{3} (1986) 267

\bibitem{pen72}R. Penrose, \emph{Techniques of differential topology in relativity} (SIAM, 1972)
\end{thebibliography}
\end{document}